\documentclass[a4paper]{aa}
\usepackage{txfonts,graphicx,natbib}
\bibpunct{(}{)}{;}{a}{}{,}
\newcommand{\pq}{\ensuremath{P_Q}}
\newcommand{\pu}{\ensuremath{P_U}}
\newcommand{\pqo}{\ensuremath{P_Q^\mathrm{(P+C)}}}
\newcommand{\pqP}{\ensuremath{P_Q^\mathrm{(P)}}}
\newcommand{\pqC}{\ensuremath{P_Q^\mathrm{(C)}}}
\newcommand{\qP}{\ensuremath{Q^\mathrm{(P)}}}
\newcommand{\qC}{\ensuremath{Q^\mathrm{(C)}}}
\newcommand{\IP}{\ensuremath{I^\mathrm{(P)}}}
\newcommand{\IC}{\ensuremath{I^\mathrm{(C)}}}
\newcommand{\Eris}{136199 Eris}
\newcommand{\Pluto}{Pluto}
\newcommand{\Charon}{Charon}
\newcommand{\Quaoar}{50000 Quaoar}
\newcommand{\Ixion}{28978 Ixion}
\newcommand{\Huya}{38628 Huya}
\newcommand{\Varuna}{20000 Varuna}
\newcommand{\DEnine}{26375 (1999 DE$_9$)}
\newcommand{\TDten}{29981 (1999 TD$_{10}$)}
\newcommand{\ELsixtyone}{136108 Haumea}

\begin{document}
\title{Discovery of two distinct polarimetric behaviours of
       trans-Neptunian objects
       \thanks{Based on observations made with ESO Telescopes at the
               La Silla-Paranal Observatory under programme ID 178.C-0036
               (PI: A.\ Barucci)}}
       \author{
        S.~Bagnulo      \inst{1}
       \and
        I.~Belskaya     \inst{2}
       \and
        K.~Muinonen     \inst{3}
       \and
        G.P.~Tozzi      \inst{4}
       \and
        M.A.~Barucci    \inst{5}
       \and
        L.~Kolokolova   \inst{6}
       \and
        S.~Fornasier    \inst{5}
        }
%\offprints{I.~Belskaya}
\institute{
           Armagh Observatory,
           College Hill,
           Armagh BT61 9DG,
           Northern Ireland, U.K.
           \email{sba@arm.ac.uk}
           \and
           Astronomical Observatory of Kharkiv National University,
           35 Sumska str., 61022 Kharkiv, Ukraine.\\
           \email{irina@astron.kharkov.ua}
           \and
           Observatory, PO Box 14, 00014 University of Helsinki, Finland.
           \email{muinonen@cc.helsinki.fi}
           \and
           INAF - Oss. Astrofisico di Arcetri,
           Largo E. Fermi 5, I-50125 Firenze, Italy.
           \email{tozzi@arcetri.astro.it}
           \and
           LESIA, Observatoire de Paris, 5,
           pl.~Jules Janssen, FR-92195 Meudon cedex, France.\\
           \email{antonella.barucci@obspm.fr, sonia.fornasier@obspm.fr}
           \and
           University of Maryland, College Park, MD, USA.
           \email{ludmilla@astro.umd.edu}
}
\authorrunning{S.\ Bagnulo et al.}
\titlerunning{Discovery of two distinct polarimetric behaviours of
trans-Neptunian Objects}

\date{Received: 2008-09-08 / Accepted: 2008-10-10}
\abstract
% Context
{
Trans-Neptunian objects (TNOs) contain the most primitive and thermally
least-processed materials from the early accretional phase of the
solar system. They allow us to study interrelations between various
classes of small bodies, their origin and evolution.
}
%Aims
{
We exploit the use of polarimetric techniques as a remote-sensing tool
to characterize the surface of TNOs.
}
%Method
{ 
Using FORS1 of the ESO VLT, we have obtained linear-polarization
measurements in the Bessell $R$ filter for five TNOs at different
values of their phase angle (i.e., the angle between the Sun, the
object, and the Earth).  Due to the large distance of the targets
($\ga 30$\,AU), the observed range of phase angles is limited to about
$0\degr-2\degr$.
}
%Results
{
We have analyzed our new observations of five TNOs, and those of
another four TNOs obtained in previous works, and discovered that
there exist two classes of objects that exhibit different
polarimetric behaviour. Objects with a diameter $> 1000\,$\,km, such
as, e.g., Pluto and Eris, show a small polarization in the scattering
plane ($\sim 0.5$\,\%) which slowly changes in the observed phase angle
range. In smaller objects such as, e.g., Ixion and Varuna, linear
polarization changes rapidly with the phase angle, and reaches $\sim
1$\,\% (in the scattering plane) at phase angle 1\degr. The
larger objects have a higher albedo than the smaller ones, and have
the capability of retaining volatiles such as CO, N$_2$ and CH$_4$. Both
of these facts can be linked to their different polarimetric
behaviour compared to smaller objects.
}
%Conclusions
{
In spite of the very limited range of observable phase angles,
ground-based polarimetric observations are a powerful
tool to identify different properties of the surfaces of TNOs. We
suggest that a single polarimetric observation at phase angle $\sim
1\degr$ allows one to determine whether the target albedo is low or
high.
}


\maketitle
%________________________________________________________________
\section{Introduction}
For a long time, measurements of linear polarization have been
exploited for the study and classification of the surfaces of small
solar-system bodies e.g., particle size, complex refractive index,
porosity, and heterogeneity are linked to the mechanism of
light scattering.

A particularly interesting phenomenon is observed at small phase
angles (the phase angle is the angle between the Sun, the object, and
the Earth), where solar-system objects like comets, asteroids,
satellites of major planets, and trans-Neptunian objects (TNOs)
exhibit \textit{negative polarization}. This is a peculiar case of
partially linearly polarized scattered light where the electric field
vector component parallel to the scattering plane predominates over
the perpendicular component, in contrast to what is expected from the
simple single Rayleigh-scattering or Fresnel-reflection
model. Negative linear polarization was first discovered by
\citet{Lyot29} in lunar observations, and later found to be a
ubiquitous phenomenon for planetary surfaces at small phase
angles. There are several physical mechanisms that explain the
existence of negative polarization from scattering of light by
particulate media, among which coherent backscattering
\citep[e.g.][]{Muinonen04} is the most relevant one for the
interpretation of TNO observations.
%%%%%%%%%%%%%%%%%%%%%%%%%%%%%%%%%%%%%%%%%%%%%%%%%%%%%%%%%%%%%%%%%%
\begin{table*}
\caption{\label{Tab_Observations}
Polarimetry of five trans-Neptunian objects in Bessell $R$ band. $\pq$
and $\pu$ are the Stokes parameters transformed such that $\pq$
is the flux perpendicular to the plane Sun-Object-Earth (the
scattering plane) minus the flux parallel to that plane, divided by
the sum of the two fluxes.}
\begin{center}
\begin{tabular}{ccrlcr@{\,$\pm$\,}lr@{\,$\pm$\,}l}
\hline \hline
Date                             &
Time (UT)                        &
\multicolumn{1}{c}{Exp}          &
\multicolumn{1}{c}{Object}       &
Phase angle                      &
\multicolumn{2}{c}{\pq}          &
\multicolumn{2}{c}{\pu}         \\
\multicolumn{1}{c}{(yyyy mm dd)} &
\multicolumn{1}{c}{(hh:mm)}      &
(sec)                            &
\multicolumn{1}{c}{}             &
\multicolumn{1}{c}{(DEG)}        &
\multicolumn{2}{c}{(\%)}         &
\multicolumn{2}{c}{(\%)}        \\
\hline
            &       &      &          &       & \multicolumn{4}{c}{}     \\
%           |       |      |          |       |         |      |         |      |
 2008 01 09 & 06:07 & 6000 & \DEnine\ & 1.412 & $-$1.39 & 0.12 &    0.00 & 0.12 \\
 2008 03 09 & 06:07 & 6000 &          & 0.110 & $-$0.20 & 0.11 &    0.08 & 0.11 \\
 2008 03 29 & 03:19 & 6400 &          & 0.519 & $-$0.66 & 0.11 & $-$0.16 & 0.11 \\[2mm]
%           |       |      |          |       |         |      |         |      |
 2008 07 08 & 09:20 & 2880 & \Eris\   & 0.600 & $-$0.11 & 0.05 &    0.04 & 0.05 \\
 2008 09 07 & 03:40 & 2880 &          & 0.379 &    0.03 & 0.05 &    0.12 & 0.05 \\[2mm]
%           |       |      |          |       |         |      |         |      |
 2007 05 10 & 01:54 & 3760 & \Huya\   & 0.614 & $-$0.73 & 0.07 & $-$0.04 & 0.07 \\
 2007 05 18 & 04:19 & 3760 &          & 0.831 & $-$0.58 & 0.07 &    0.08 & 0.07 \\
 2007 07 17 & 01:44 & 3760 &          & 1.984 & $-$1.61 & 0.07 &    0.00 & 0.07 \\
 2008 03 05 & 07:40 & 6000 &          & 1.624 & $-$1.27 & 0.06 &    0.05 & 0.06 \\
 2008 05 30 & 04:54 & 5120 &          & 1.115 & $-$1.10 & 0.06 &    0.04 & 0.06 \\[2mm]
%           |       |      |          |       |         |      |         |      |
 2006 11 26 & 07:05 & 5800 & \Varuna\ & 0.91  & $-$1.04 & 0.12 &    0.08 & 0.11 \\
 2006 12 14 & 06:03 & 5800 &          & 0.572 & $-$0.45 & 0.15 & $-$0.11 & 0.14 \\
 2007 01 13 & 02:55 & 5800 &          & 0.135 & $-$0.22 & 0.10 &    0.04 & 0.10 \\
 2008 03 29 & 01:10 & 7200 &          & 1.301 & $-$1.18 & 0.13 &    0.06 & 0.13 \\[2mm]
%           |       |      |          |       |         |      |         |      |
 2008 08 09 & 23:59 &  960 &\ELsixtyone\
                                      & 0.987 & $-$0.68 & 0.06 & $-$0.02 & 0.06 \\[2mm]
\hline
\end{tabular}
\end{center}
\end{table*}
%%%%%%%%%%%%%%%%%%%%%%%%%%%%%%%%%

For many solar-system objects, it is possible to measure the behaviour
of the polarization for an extended range of phase angles, and
identify at least three important characteristics, i.e.:
\textit{i)} the slope of the polarimetric curve; \textit{ii)} the
minimum polarization; \textit{iii)} the inversion angle at which the
polarization changes from being parallel to the scattering plane and
becomes, at larger phase angle values, perpendicular to the scattering
plane. In contrast, ground-based polarimetric observations of TNOs can
cover only a very limited phase-angle range (due to the large distance
of these objects). The observed polarized light is always parallel to
the scattering plane, and it is not possible, in general, to
estimate the polarization minimum.

The first polarimetric observations of a TNO (except Pluto) were
carried out by \citet{Boeetal04} for \Ixion. Within the very
narrow range of the observed phase angles ($0.2\degr-1.3\degr$), they
revealed a pronounced negative polarization changing rapidly as a
function of the phase angle. Since then, three other TNOs have been
the subject of a detailed study: \TDten\ \citep{Rousetal05},
\Quaoar\ \citep{Bagetal06}, and \Eris\
\citep{Beletal08}. \citet{Beletal08} noted that the available
polarimetric data for TNOs show negative polarization with two
different trends at small phase angles. Only a small fraction of
linear polarization is measured in the largest objects \Pluto, \Eris,
and \Quaoar, with only subtle changes as a function of phase angle.  This
is in contrast with the steep gradient of about $-1$\,\% per degree that
was measured for \Ixion.

In this paper we present 13 observations of another four
TNOs, and two new observations of \Eris. These
new data allow us to generalize the finding by \citet{Beletal08}.

\section{Observations}
Fifteen new broadband linear polarization measurements of five TNOs
were obtained from November 2006 to September 2008 at the ESO
Very Large Telescope (VLT) with FORS1 \citep{Appetal98}.  Using the
$R$ Bessell filter, we obtained two new measurements of Eris, that was
already observed by \citet{Beletal08}, and thirteen measurements of
\DEnine, \Huya, \ELsixtyone\ (2003 EL$_{61}$), and \Varuna, that were
never before observed in polarimetric mode.

The heliocentric distance of the observed TNOs is $\ga 30$\,AU. As a
consequence, the observable phase-angle range is very limited,
compared to what can be achieved for most asteroids and comets. The
phase angle range that was sampled is $\la 2\degr$
for all newly observed TNOs.

We aimed at obtaining polarization measurements with error bars
between 0.05\,\% and 0.1\,\%, which requires a signal-to-noise ratio
between 1000 and 2000 (cumulated on both beams and all positions of
the retarder waveplate). Since our targets are faint ($R\sim 18-20$),
these observations were possible only by using an 8-m telescope. In
order to optimize the phase-angle sampling, our observations were
scheduled in service mode. Observing blocks were planned so as to avoid
too bright a background due to lunar illumination, and to avoid epochs
when targets were too close to bright stars.

Polarimetric observations were generally performed with the retarder
waveplate at all positions between 0\degr\ and 157.5\degr, at
22.5\degr\ steps. For each observation, the exposure time cumulated
over all exposures varied from 24 minutes (for \Eris) to
2\,h (for \Varuna).  Raw data were then treated as explained in
\citet{Bagetal06}, and our measurements are reported adopting as a
reference direction the perpendicular to the great circle
passing through the object and the Sun. This way, $\pq$ represents
the flux perpendicular to the plane Sun-Object-Earth (the scattering
plane) minus the flux parallel to that plane, divided by the
sum of these fluxes. For symmetry reasons, $\pu$ values are always
expected to be zero, and inspection of their values allows us
to perform an indirect quality check of the \pq\ values.
%%%%%%%%%%%%%%%%%%%%%%%%%%%%%%%%%%%%%%%%%%%%%%%%%%%%%%%%%%%%%%%%%%%%%%%%%
\begin{figure}
\begin{center}
\scalebox{0.37}{
\includegraphics*[0cm,5.8cm][22cm,25cm]{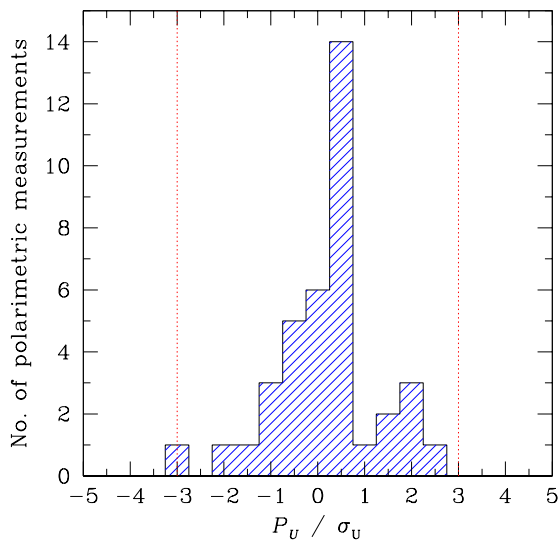}}
\end{center}
\caption{\label{Fig_Histo_Pu} Distribution of the \pu\ values normalized
to their error bars. The null detection of \pu\ values serves as a quality check
for the \pq\ measurements of Fig.~\ref{Fig_Pq_Data}.
}
\end{figure}
%%%%%%%%%%%%%%%%%%%%%%%%%%%%%%%%%%%%%%%%%%%%%%%%%%%%%%%%%%%%%%%%%%%%%%%%%

Our new polarimetric measurements are given in
Table~\ref{Tab_Observations}.  These data have enlarged the number of
TNOs for which polarimetric observations are available by a factor of
two. Previously published data, that will be also considered in our
analysis, include seven measurements of \Pluto\
\citep{BreCoc82}, nine measurements of \Ixion\
\citep{Boeetal04}, five measurements of \TDten\ \citep{Rousetal05},
five measurements of \Quaoar\ \citep{Bagetal06}, and four measurements
of \Eris\ \citep{Beletal08}.  For the sake of consistency, data of
\TDten\ were re-reduced adopting exactly the same reduction procedure
as for the remaining observations obtained with FORS1, leading to values
only slightly different from the previously published ones.

All data considered here, except for \Pluto, were obtained with the
FORS1 instrument with the Bessell $R$ filter.  Pluto's polarimetry
refers to a filter similar to Bessell $V$.  We note that
\citet{Bagetal06} obtained polarimetric measurements of the Centaur
Chiron in the  Bessell $B$, $V$, and $R$ bands at six different phase
angles. At each phase angle, the polarimetric measurements obtained in
the three different bands appear relatively consistent among
themselves. A similar behaviour was found for \TDten\ that was
observed both in the Bessell $R$ and $V$ filters by
\citet{Rousetal05}. Additional measurements of Pluto obtained by
\citet{KelFix73} and \citet{Avretal92} with no filter are also consistent
with those by \citet{BreCoc82}. This suggests that even though \Pluto\
polarimetry was obtained in a different band than the other TNOs, its
comparison with new data obtained in the $R$ band is still meaningful.

It should also be recalled that Pluto is in fact a double system.
Yet, in the observed phase-angle interval, Pluto's intrinsic
polarization is confined within the range $-0.35$\,\% and $-0.1$\,\%,
for the following reason. We denote with $Q$ the Stokes parameter not
normalised to the intensity $I$, and with $\pqP = \qP/\IP$ and
$\pqC=\qC/\IC$ \Pluto's and \Charon's intrinsic polarization,
respectively. We can assume that, at each phase angle, Pluto, Charon,
the Sun, and the Earth define an identical scattering plane, so that
\pqP\ and \pqC\ are expressed in the same reference system. This
allows us to write, for the observed polarization of the double
system, $\pqo = (\qP + \qC) / (\IP + \IC)$. Taking into account that
the ratio between the reflected light of the two objects is $\sim
0.17$, we deduce that
%%%%%%%%%%%%%%%%%%%%%%%%%%%%%%%%%%%%
\begin{equation}
\pqP = \pqo + 0.17 \, (\pqo - \pqC) \; . \label{Eq_Pluto}
\end{equation}
%%%%%%%%%%%%%%%%%%%%%%%%%%%%%%%%%%%%
Assuming that \Charon\ has $-1.5\,\%\la \pqC \la 0\,\%$, and
considering for the total polarization its mean value of $-0.3$\,\%,
Eq.~(\ref{Eq_Pluto}) tells us that \pqP\ ranges between $-0.35$\,\%
(in the case of $\pqC = 0$) and $\simeq -0.1$\,\% (in the case of
$\pqC = -1.5$\,\%).

Some polarimetric measurements are close to the limit of instrumental
polarization, which in FORS1 is $\la 0.04$\,\%
\citep{Fosetal07}. Instrumental polarization is difficult to subtract
from science data, since it depends on the instrument setting and
telescope orientation but, for the same reason, we can assume
that instrumental polarization does not introduce any
\textit{systematic} offset. An indirect
confirmation that instrumental polarization does not introduce systematic
offsets comes from inspection of the measured \pu\
values. Figure~\ref{Fig_Histo_Pu} shows the distributions of the \pu\
values for all objects observed with FORS1 expressed in error bar
units (for most of new data between 0.05\,\% and 0.12\,\%). Due to the
relatively limited number of measurements (38) we cannot expect to
reproduce a Gaussian distribution.  Yet, the fact that the \pu\
distribution is roughly centered at zero, and all points are within
$-3\ \le \pu/\sigma_U \le 3$, fully supports the reliability of the
polarimetric measurements.

Figure~\ref{Fig_Pq_Data} shows the \pq\ values as a function of the
phase angle measured for all data reported in
Table~\ref{Tab_Observations}, and those previously published listed
above.  The left panel of Fig.~\ref{Fig_Pq_Data} refers to the larger TNOs
\Eris, \Pluto, \Quaoar, and \ELsixtyone. The right panel shows the
polarization phase angle dependence for the remaining objects: \Ixion, \Huya,
\DEnine, \TDten, and \Varuna.

\section{Results and discussion}
%%%%%%%%%%%%%%%%%%%%%%%%%%%%%%%%%%%%%%%%%%%%%%%%%%%%%%%%%%%%%%%%%%%%%%%%%
\begin{figure*}
\begin{center}
\rotatebox{270}{
\scalebox{0.51}{
\includegraphics*[3.6cm,0cm][18.6cm,27cm]{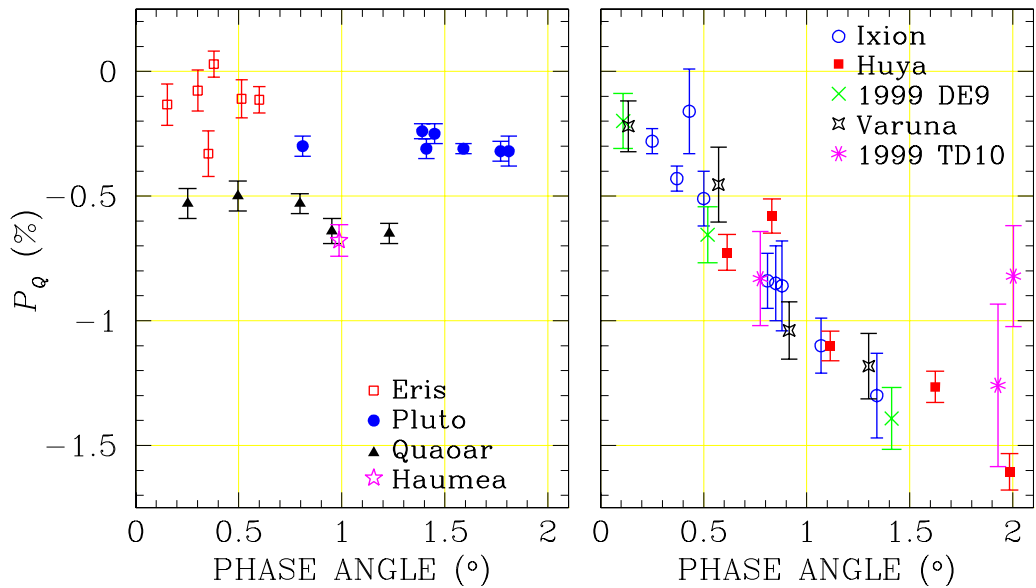}}}
\end{center}
\caption{\label{Fig_Pq_Data}
Linear-polarization measurements of nine TNOs as a function of the phase angle.
}
\end{figure*}
%%%%%%%%%%%%%%%%%%%%%%%%%%%%%%%%%%%%%%%%%%%%%%%%%%%%%%%%%%%%%%%%%%%%%%%%%
Figure~\ref{Fig_Pq_Data} shows that larger TNOs exhibit a small
fraction of negative linear polarization roughly constant in the
observed phase angle range.  As far as \ELsixtyone\ is concerned, the
only conclusion that can be drawn from its single measurement is that
the polarization is very similar to that measured for Quaoar and
higher (in absolute value) than that of \Pluto\ and \Eris. Whether its
phase angle dependence resembles that of other large objects will have
to be checked with new observations. Our new polarimetric measurements
of Eris, which were obtained with a higher accuracy than before,
confirmed our previous findings about its small negative polarization
\citep{Beletal08} and expanded the observed phase angle range to the
maximum range presently reachable for this distant object.  In
particular, a previous observation obtained at the phase angle of
0.35\degr\ showed the strongest negative polarization for this object
\citep[$\sim -0.3$\,\%, see][]{Beletal08}, whereas our new measurement
obtained at a similar phase angle is practically consistent
with zero.  Therefore we cannot confirm the presence of a negative 
polarization surge around that phase angle.

The new observations of the classical TNOs \Varuna, \Huya\ and the
scattered-disk object \DEnine\ reveal a pronounced negative
polarization changing rapidly with phase angle and reaching about
$-1$\,\% at the phase angle of 1\degr. Similar polarization behaviour
was previously found for the resonant object \Ixion\ by
\citet{Boeetal04}, who pointed out that this object exhibits the most
pronounced negative polarization measured for a solar-system body so
far, and raised the question of whether this was a unique case or typical
for TNOs. Our new observations have proved that the high negative
polarization at small phase angles is quite typical for TNOs, as three
newly observed objects have shown a polarimetric behaviour similar to
that of Ixion.  In fact, all four ``small'' TNOs together show a
strikingly similar polarimetric behaviour that is practically
indistinguishable within the error bars.

Data for \TDten\ \citep{Rousetal05} extend up to phase angle $\sim
3\degr$. In the phase-angle range $0\degr-2\degr$, the observed
polarization behaviour is consistent with those of the other small
objects, but this should be confirmed by a more refined sampling of
this range.  At phase angles 2\degr\ and 3\degr\ the observed
polarization is about $-1$\,\%, which suggests a polarization minimum
in that range.  Observations of Centaurs Chiron obtained at larger
phase angles compared to TNOs reveal a polarization minimum in the
range $\sim 1.5\degr - 2\degr$ \citep{Bagetal06}, and new unpublished
observations suggest a similar behaviour for Centaur Pholus (Belskaya
et al., in preparation). All this leads us to speculate that the
minimum of polarization of TNO phase curves are at phase angles
slightly larger than 1.5\degr, perhaps between 1.5\degr\ and 3\degr,
which would be noticeably different from that of asteroids and comets,
that show a minimum between 7\degr\ and 10\degr\ \citep[see,
e.g.,][]{Penetal05}.

In spite of the very limited observed phase angle range, polarimetric
observations of TNOs reveal two different behaviours.  TNOs with a
diameter smaller than 1000\,km exhibit a negative polarization that
rapidly increases (in absolute value) with the phase angle, reaching
about $1$\,\% at the phase angle of 1\degr.  Larger TNOs exhibit a
small fraction of negative linear polarization ($\la 0.7$\,\%)
which does not noticeably change in the observed phase angle range.

It is quite natural to associate the two different behaviours of
polarization phase dependencies with a different composition and/or
structure of the surfaces of the objects. The two groups of objects
with different polarimetric properties differ not only in size but
also in surface albedo, which is higher in the larger
objects than in the smaller ones. Although uncertainties in TNO albedo
determination are quite large, it is evident that darker surfaces
exhibit higher negative polarization than brighter surfaces. This
trend resembles the dependence of the polarization minimum on the albedo
found for asteroids, but the lack of accurate albedo data prevents
us from attempting to obtain firm relationships between albedo and
polarization such as those obtained for asteroids \citep[e.g.,][]{LupMoh96}.
Yet, even a single measurement of linear polarization of a TNO at
phase angle $\sim 1\degr$ can provide at least a
distinction between high- and low-albedo surfaces.

At first glance, our results seem contradictory to the predictions of
the coherent-backscattering mechanism considered to be the most
probable cause of negative polarization at small phase angles. The
coherent-backscattering mechanism results in a sharp surge of negative
polarization accompanied by narrow brightness opposition peaks that
should be more prominent for the brighter surfaces. Such surges were
found for bright satellites and asteroids with a peak polarization of
about $-0.4$\,\% centered at the phase angle of $0.2\degr-1\degr$
\citep[for a review, see][] {Misetal06}. The observations of Eris and
Pluto do not show opposition surges in polarization or brightness in
the phase-angle ranges covered (down to 0.15\degr\ and 0.80\degr,
respectively).  The observations can, however, be well explained by a
narrower width of the coherent-backscattering opposition surges for
large TNOs as compared to those observed for satellites and asteroids
\citep[see discussion in][]{Beletal08}. Note that, according to
laboratory measurements, some bright powdered samples also do not show
a negative-polarization surge in the phase-angle range of 0.2-4 deg
\cite[see][]{Shketal02}.

Perhaps the most important difference between the surface
characteristics of the objects that exhibit a different polarimetric
behaviour is that the TNOs with small and constant negative
polarization are supposed to have the capability of retaining
volatiles such as CO, N$_2$ and CH$_4$ \citep{SchaBro07}. \Eris\ and
\Pluto\ have methane rich surfaces. The other two objects \ELsixtyone\
and \Quaoar, with slightly higher polarization (in absolute value),
exhibit dominantly water ice spectra, and are believed to be in a
transition phase where not all volatiles are lost yet
\citep{Brown08}. This evolutionary phase may explain their
different albedo, as well as their different polarimetric behaviour
compared to smaller objects, which have certainly lost the volatile
components.


\begin{thebibliography}{}
\bibitem[Appenzeller et al.(1998)]{Appetal98} Appenzeller, I., Fricke, K.,
                          Furtig, W., et al. 1998, The Messenger, 94, 1
\bibitem[Avramchuk et al.(1992)]{Avretal92} 
                          Avramchuk, V.~V., Rakhimov, V.~I., Chernova, G.~P., \&
                          Shavlovskii, N.~M.
                          1992, Kinemat.\ Fiz.\ Nebesn. Tel, 8, N4, 37.
\bibitem[Bagnulo et al.(2006)]{Bagetal06}
                          Bagnulo, S., Boehnhardt, H., Muinonen, K., et al.
                          2006, A\&A, 450, 1239
\bibitem[Belskaya et al.(2008)]{Beletal08}
                          Belskaya, I., Bagnulo, S., Muinonen, K., et al.
                          2008, A\&A, 479, 265
\bibitem[Boehnhardt et al.(2004)]{Boeetal04}
                          Boehnhardt, H., Bagnulo, S., Muinonen, K., et al.
                          2004, A\&A, 415, L21
\bibitem[Breger \& Cochran(1982)]{BreCoc82}
                          Breger, M., \& Cochran, W.~D. 1982, Icarus, 49, 120
\bibitem[Brown(2008)]{Brown08}
                          Brown, M.E. 2008, in:
                          The Solar System beyond Neptune. Barucci A.M.,
                          Boehnhardt, H., Cruikshank, D.P., \& Morbidelli, A.
                          (eds.), Univ. Arizona Press, p.\ 335
\bibitem[Fossati et al.(2007)]{Fosetal07}
                          Fossati, L., Bagnulo, S., Mason, E., \&
                          Landi Degl'Innocenti, E. 2007, in: C. Sterken                          (ed.), ASP Conference Series, vol. 364, p.\,503
\bibitem[Kelsey \& Fix(1973)]{KelFix73}
                          Kelsey, J.~D., \& Fix, L.~A. 1973, AJ, 184, 633
\bibitem[Lupishko \& Mohamed(1996)]{LupMoh96} 
                           Lupishko D.F., \& Mohamed R.A. 1996, Icarus, 11, 209
\bibitem[Lyot(1929)]{Lyot29}
                           Lyot, B. 1929, Ann.\ Obs.\ Paris, vol.~8, 1
\bibitem[Mishchenko et al.(2006)]{Misetal06}
                          Mishchenko, M.~I.,Rosenbush, V.~K., \& Kiselev, N.~N.
                          2006, Applied optics, 45, 4459
\bibitem[Muinonen(2004)]{Muinonen04}
                          Muinonen, K.
                          2004, Waves in Random Media, vol.~14, Iss.~3, p.~365
\bibitem[Penttil\"{a} et al.(2005)]{Penetal05}
                          Penttil\"{a}, A., Lumme, K., Hadamcik, E., 
                          Levasseur-Regourd, A.C. 2005,
                          A\&A, 432, 1081
\bibitem[Rousselot et al.(2005)]{Rousetal05}
                          Rousselot P., Levasseur-Regourd A.C., Muinonen K., \&  Petit
                          J.-M. 2005, Earth, Moon, \& Planets, 97, 353
\bibitem[Schaller \& Brown(2008)]{SchaBro07}
                          Schaller, E.L., \& Brown, M.E. 2008, ApJ, 659, L61
\bibitem[Shkuratov et al.(2002)]{Shketal02} Shkuratov Yu., Ovcharenko A.,
                          Zubko E., et al. 2002, Icarus, 159, 396
\end{thebibliography}
\end{document}